\documentclass[aps,prl,twocolumn,superscriptaddress,showpacs,preprintnumbers,amsmath,amssymb,floatfix,nofootinbib]{revtex4}

\usepackage{graphicx} 
\usepackage{dcolumn}  
\usepackage{rotating} 
\usepackage{lscape}

\usepackage{bm}
\usepackage{url} 
\usepackage{amsmath} 
\usepackage{amssymb} 

\usepackage{verbatim}
\usepackage{multirow}

\graphicspath{{ps}}

\begin{document}

\title{First axion dark matter search with toroidal geometry}
\author{J. Choi}\affiliation{Center for Axion and Precision Physics
  Research (CAPP), Institute for Basic Science (IBS), Daejeon 34141, Republic
of Korea}
\author{H. Themann}\affiliation{Center for Axion and Precision Physics
  Research (CAPP), Institute for Basic Science (IBS), Daejeon 34141, Republic
of Korea}
\author{M. J. Lee}\affiliation{Center for Axion and Precision Physics
  Research (CAPP), Institute for Basic Science (IBS), Daejeon 34141, Republic
  of Korea}
\author{B. R. Ko}\email[Corresponding author~:~]{brko@ibs.re.kr}\affiliation{Center for Axion and Precision Physics
  Research (CAPP), Institute for Basic Science (IBS), Daejeon 34141, Republic
of Korea}
\author{Y. K. Semertzidis}\affiliation{Center for Axion and Precision
  Physics Research (CAPP), Institute for Basic Science (IBS), Daejeon 34141,
  Republic of Korea}\affiliation{Dept. of Physics, Korea Advanced
  Institute of Science and Technology (KAIST), Daejeon 34141, Republic
  of Korea}

\begin{abstract}
  We firstly report an axion haloscope search with toroidal
  geometry. In this pioneering search, we exclude the axion-photon
  coupling $g_{a\gamma\gamma}$ down to about $5\times10^{-8}$
  GeV$^{-1}$ over the axion mass range from 24.7 to 29.1 $\mu$eV at a
  95\% confidence level. The prospects for axion dark matter searches
  with larger scale toroidal geometry are also considered.  
\end{abstract}


\maketitle
\tighten

{\renewcommand{\thefootnote}{\fnsymbol{footnote}}}
\setcounter{footnote}{0}

According to precision cosmological measurements~\cite{PLANCK}, more
than 80\% of the matter content in the Universe is now believed to be
cold dark matter (CDM). However, the CDM composition is beyond the
standard model of particle physics, and thus is still unknown to
date. One of the most compelling candidates for CDM is the
axion~\cite{AXION}, provided its mass is above 1
$\mu$eV~\cite{CDM_LOW} and below 3 meV~\cite{SN1987}. The axion is the
result of the breakdown of a new symmetry proposed by Peccei and
Quinn~\cite{PQ} to solve the strong $CP$ ($C$harge-conjugation and
$P$arity) problem in Quantum Chromodynamics~\cite{strongCP}. As a
result of the axion production mechanism, the axion mass range is very
large with the above range being optimum for CDM. 

The axion search method proposed by Sikivie~\cite{sikivie}, also known
as the axion haloscope search, involves a microwave resonant cavity
with a static magnetic field that induces axion conversions into
microwave photons. The conversion power corresponding to the axion
signal should be enhanced when the axion mass $m_a$ matches the
resonant frequency of the cavity mode $\nu$, $m_a=h\nu/c^2$. The power
would be detected as the axion signal is
\begin{equation}
  P_{a}=g^2_{a\gamma\gamma}\frac{\rho_a\hbar^2}{m^2_a c}\omega 2U_M C\frac{Q\beta}{(1+\beta)^2},
  \label{EQ:PAXION}
\end{equation}
where $g_{a\gamma\gamma}$ is the axion-photon coupling strength, whose
two popular benchmark models are KSVZ~\cite{KSVZ} for hadronic axions
and DFSZ~\cite{DFSZ}, which also includes axion coupling to leptons,
$\rho_a\approx0.45$ GeV/cm$^3$ is the local dark matter density,
$\omega=2\pi\nu$, and $U_M=\frac{1}{2\mu_0}B^2_{\rm
  avg}V\equiv\frac{1}{2\mu_0}\int\vec{B}^2 dV$ is energy stored in a
magnetic field in the cavity volume $V$, where $\vec{B}$ is a static
magnetic field provided by magnets in the axion haloscopes. The
cavity-mode-dependent form factor $C$ whose general definition can be
found in Ref.~\cite{EMFF_BRKO} and quality factor $Q$ are also shown in
Eq.~(\ref{EQ:PAXION}) and $Q/(1+\beta)$ corresponds to the loaded
quality factor $Q_L$, where $\beta$ denotes the mode coupling to the
load. Assuming the axions have an isothermal distribution, the signal
power given in Eq.~(\ref{EQ:PAXION}) would then distribute over a
Maxwellian shape with an axion rms speed of about 270 km/s in our
galaxy~\cite{AXION_SHAPE}, which is the basic model considered in this
paper.

Most of the axion haloscope searches to
date~\cite{haloscope2,haloscope3,haloscope4,haloscope5,HF} have
employed a cylindrical resonator without an open
resonator~\cite{Orpheus}. In this paper, we report the first axion
haloscope search with toroidal geometry. As reported in our previous
publication~\cite{EMFF_BRKO}, as long as
$\vec{\nabla}\times{\vec{B}}\sim0$ is valid, Eq.~(\ref{EQ:PAXION}) and
the definition of the cavity-mode-dependent form factor $C$ therein
are valid as they are in axion haloscopes in spite of the lack of an
axion to a photon magnetic field coupling in the form factor $C$,
which is also true for toroidal geometry.
We refer to this axion dark matter search as {\bf ACTION} for
``{\bf A}xion haloscopes at {\bf C}APP with {\bf T}oro{\bf I}dal res{\bf ON}ators''.
The prospects for larger scale ACTION experiments are also discussed.

\begin{figure}[htbp]
  \centering
  \includegraphics[height=0.23\textwidth]{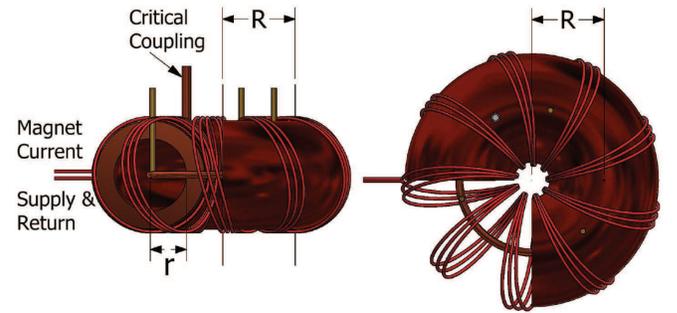}  
  \caption{Lateral (left) and top (right) views of the toroidal cavity,
    frequency tuning system, and toroidal coils whose dimensions are
    given in the text. Note that it is a cut-away view to show up
    details of the system.}
  \label{FIG:CAPPUCCINO}
\end{figure}

The ACTION experiment considered in this paper constitutes a tunable
copper toroidal cavity, toroidal coils which provide a static magnetic
field, and a typical heterodyne receiver chain. The experiment was
conducted at room temperature. The system is called the ``simplified
ACTION''. A torus is defined by $x=(R+r\cos\theta)\cos\phi$,
$y=(R+r\cos\theta)\sin\phi$, and $z=r\sin\theta$, where $\phi$ and
$\theta$ are angles that make a full circle of radius $R$ and $r$,
respectively. As shown in Fig.~\ref{FIG:CAPPUCCINO}, $R$ is the
distance from the center of the torus to the center of the tube and
$r$ is the radius of the tube. Our cavity tube's $R$ and $r$ are 4 and
2 cm, respectively, and the cavity thickness is 1 cm.
The frequency tuning system constitutes a copper tuning hoop whose $R$
and $r$ are 4 and 0.2 cm, respectively, and three brass posts for
linking between the hoop and a piezo linear actuator that controls the
movement of our frequency tuning system.
The quasi-TM$_{010}$ (QTM$_1$) modes of the cavity are tuned by moving
up and down our frequency tuning system along the axis parallel to the
brass posts. Two magnetic loop couplings were employed, one for weakly
coupled magnetic loop coupling and the other for critically coupled
magnetic loop coupling, i.e. $\beta\simeq1$ to maximize the signal
power expressed in Eq.~(\ref{EQ:PAXION}). The $Q_L$ values of the
QTM$_1$ modes with $\beta\simeq1$ vary from $\sim$10,000 at the $\nu$
of 6 GHz to $\sim$5,000 at that of 7 GHz~\cite{WHYNOT8GHz}, where the
former and the latter correspond to about 95\% and 67\% of the
designed values. Without the frequency tuning system, the $Q_L$ of our
toroidal cavity is almost 100\% of the designed $Q_L$, which is
attributed to the absence of the cavity endcaps in toroidal
geometry. The rather lower $Q_L$ than that of the designed $Q_L$ could
be improved by optimizing the magnetic loop coupling positions. We
could not measure the $Q_L$ of the QTM$_1$ confidently for the $\nu$
from 6.77 to 6.80 GHz due to mode-crossings at the $\nu$ of about 6.78
GHz. Based on the $Q_L$ values of the QTM$_1$ modes, the discrete
frequency step was chosen to be a half of the smallest cavity
linewidth, i.e., 300 kHz.

\begin{figure}[htbp]
  \centering
  \includegraphics[height=0.32\textwidth,width=0.5\textwidth]{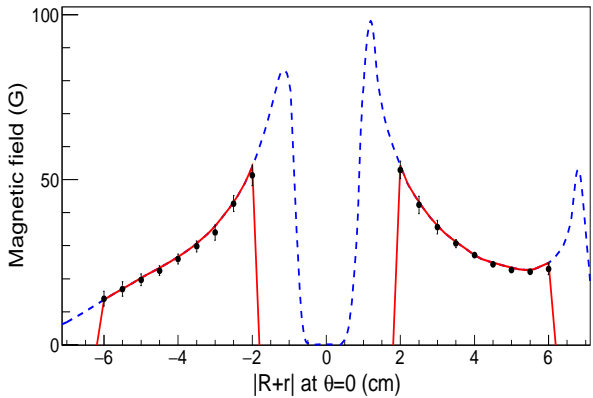}
  \caption{Magnetic field as a function of radial position $|R+r|$ at
    $\theta=0$. Dashed (blue) lines are obtained from the finite
    element method and correspond to the toroidal cavity system, and
    solid lines (red) correspond to the cavity tube. Dots with error
    bars are measurement values. The results at positive $R+r$ are
    along a coil, while those at negative $R+r$ are between two
    neighboring coils.}
    \label{FIG:BSIMULATION}
\end{figure}
A static magnetic field was provided by a 1.6 mm diameter copper wire
ramped up to 20 A with three winding turns, as shown in
Fig.~\ref{FIG:CAPPUCCINO}. Figure~\ref{FIG:BSIMULATION} shows good
agreement between measurement with a Hall probe and a
simulation~\cite{CST} of the magnetic field induced by the coils. The
$B_{\rm avg}$ from the magnetic field map provided by the simulation
turns out to be 32 G. No fringe magnetic fields are outside of an
ideal toroidal magnet, and as shown in Fig.~\ref{FIG:BSIMULATION}, the
magnetic field outside of our toroidal magnet system also drops
drastically, as expected. Therefore, the handling of
quantum-noise-limited superconducting amplifiers that work in an
almost zero field environment is much easier with toroidal geometry.

\begin{figure}[htbp]
  \centering
  \includegraphics[height=0.3\textwidth,width=0.5\textwidth]{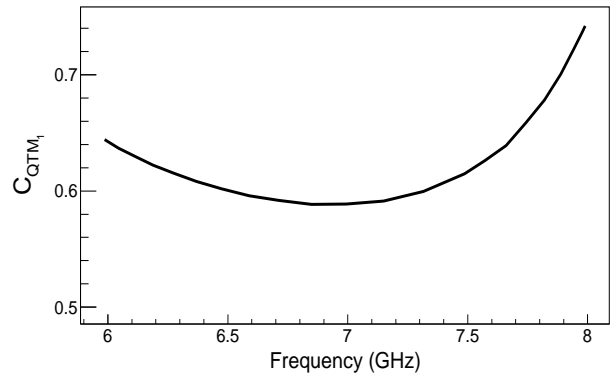}
  \caption{Form factors of the QTM$_1$ mode of the toroidal cavity as a
    function of the QTM$_1$ frequency.}
  \label{FIG:FF}
\end{figure}

\begin{figure}[htbp]
  \centering
  \includegraphics[height=0.35\textwidth,width=0.5\textwidth]{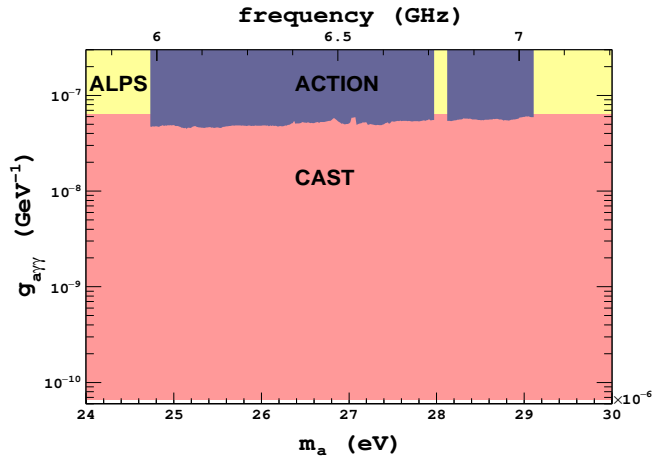}
  \caption{Excluded parameter space at the 95\% C.L. by this
    experiment together with previous results from
    ALPS~\cite{ALPS2010} and CAST~\cite{CAST2017}. No limits are set
    from 6.77 to 6.80 GHz due to with a quasi-TE mode in that
    frequency region and the TE mode is also confirmed by a
    simulation~\cite{CST}.}
  \label{FIG:LIMIT}
\end{figure}

With the magnetic field map and the electric field map of the QTM$_1$
mode in the toroidal cavity, we numerically evaluated the form factor
of the QTM$_1$ mode as a function of the QTM$_1$ frequency, as shown in
Fig.~\ref{FIG:FF}, where the highest frequency appears when the
frequency tuning system is located at the center of the cavity tube,
such as in Fig.~\ref{FIG:CAPPUCCINO}. As shown in Fig.~\ref{FIG:FF},
we found no significant drop-off in the form factors of the QTM$_1$
modes, which is attributed to the absence of the cavity endcaps in
toroidal geometry. A drop-off in the form factor happens when hybrid
modes occur due to the capacitive effect caused by the gaps between
the tuning rod and the cavity endcaps in cylindrical geometry, which
was confirmed by simulation~\cite{CST}. While we cannot avoid such
gaps in cylindrical geometry, no such gaps exist in toroidal geometry
in the absence of the cavity endcaps, as shown in
Fig.~\ref{FIG:CAPPUCCINO}. This lack of the form factor drop-off is
one of the significant advantages in toroidal geometry.
Note that the QTM$_1$ mode can be identified in a simulation
regardless of mode-crossings; thus, the associated form factor can be
obtained regardless of mode-crossings, as shown in
Fig.~\ref{FIG:FF}. On the other hand, experiments with minimal
coupling to the crossed mode cannot distinguish the QTM$_1$ mode from
the crossed mode; hence, the associated form factor is not very
meaningful experimentally in that case. In toroidal geometry, the
situation is even more favorable, with the competing mode appearing
only very weakly due to the azimuthal degeneracy of the quasi-TE
modes. Experimentally, we found out that the only interference with
the QTM$_1$ mode is located at the frequency gap between 6.77 and 6.80
GHz, as mentioned earlier (see also Fig.~\ref{FIG:LIMIT}).

Our receiver chain consists of a single data acquisition channel that
is analogous to that adopted in ADMX~\cite{ADMX_NIM} except for the
cryogenic parts. Power from the cavity goes through a directional
coupler, an isolator, an amplifier, a band-pass filter, and a mixer,
and is then measured by a spectrum analyzer at the end. Cavity
associates, $\nu$, and $Q_L$ are measured with a network analyzer by
toggling microwave switches. The gain and noise temperature of the
chain were measured to be about 35 dB and 400 K, respectively, taking
into account all the attenuation in the chain, for the frequency range
from 6 to 7 GHz.

The signal-to-noise ratio (SNR) considered in this paper is
\begin{equation}
  {\rm SNR}=\frac{P_{a,g_{a\gamma\gamma}\sim6.5\times10^{-8}~{\rm GeV}^{-1}}}{P_n}\sqrt{N},
  \label{EQ:SNR}
\end{equation}
where $P_{a,g_{a\gamma\gamma}\sim6.5\times10^{-8}~{\rm GeV}^{-1}}$ is
the signal power when $g_{a\gamma\gamma}\sim6.5\times10^{-8}$
GeV$^{-1}$, which is approximately the limit achieved by the ALPS
collaboration~\cite{ALPS2010}. $P_n$ is the noise power equating to
$k_B T_n b_a$, and $N$ is the number of power spectra, where $k_B$ is
the Boltzmann constant, $T_n$ is the system noise temperature which is
a sum of the noise temperature from the cavity ($T_{n,{\rm cavity}}$)
and the receiver chain ($T_{n,{\rm chain}}$), and $b_a$ is the signal
bandwidth. We iterated data taking as long as $\beta\simeq 1$, or
equivalently, a critical coupling was made, which resulted in about 3,500
measurements. In every measurement, we collected 3,100 power spectra and
averaged them to reach at least an SNR in Eq.~(\ref{EQ:SNR}) of about
8, which resulted in an SNR of 10 or higher after overlapping the
power spectra at the end.
\begin{table*}
\caption{Expected experimental parameters for the ACTION
  experiments. Note that it lists feasible lowest and highest search
  ranges only for single-cavity experiments. The 4-cavity experiment
  parameters are chosen to search for the unexplored region by
  single-cavity experiments. We set $T_{n,{\rm chain}}$ approximately
  equals to the temperature of the 1st amplifier in the chain, $T_{n,
    {\rm amp}}$.}
\begin{ruledtabular}
  \begin{tabular}{lcccccccccccc}
    \multirow{2}{*}{experiment}&$B_{\rm avg}$&$R$&$r$&tuning&search&\multicolumn{2}{c}{search range}&form&$Q_L$&$T_{n,{\rm cavity}}$&$T_{n,{\rm chain}}$&DFSZ $d\nu/dt$ \\
    &(T)&(cm)&(cm)&hoop&mode&($\mu$eV)&(GHz)&factor&with $\beta=2$&(K)&(K)&(GHz/year) \\ \hline
    large ACTION&\multirow{2}{*}{5}&\multirow{2}{*}{200}&\multirow{2}{*}{50}&alumina&QTM$_1$&[0.79, 0.95]&[0.19, 0.23]&0.4&120,000&\multirow{2}{*}{3.73}&\multirow{2}{*}{1.00}&0.43 \\
    (single-cavity)&&&&copper&QTM$_2$&[2.23, 2.73]&[0.54, 0.66]&0.08&111,000&&&0.13 \\ \hline
    large ACTION&\multirow{2}{*}{5}&\multirow{2}{*}{200}&\multirow{2}{*}{20}&\multirow{2}{*}{copper}&QTM$_1$&[2.48, 3.51]&[0.60, 0.85]&0.6&47,000&\multirow{2}{*}{3.73}&\multirow{2}{*}{1.00}&1.86 \\
    (4-cavity)&&&&&QTM$_2$&[5.66, 7.24]&[1.37, 1.75]&0.08&63,000&&&0.21 \\ \hline
    small ACTION&\multirow{2}{*}{12}&\multirow{2}{*}{50}&\multirow{2}{*}{9}&alumina&QTM$_1$&[4.38, 5.25]&[1.06, 1.27]&0.4&40,000&\multirow{2}{*}{0.09}&0.09&6.27 \\
    (single-cavity)&&&&copper&QTM$_2$&[12.32, 15.05]&[2.98, 3.64]&0.08&46,000&&0.23&0.74 \\     
  \end{tabular}
\end{ruledtabular}
\label{TABLE:PROSPECTS}
\end{table*}

Our overall analysis basically follows the pioneer study described in
Ref.~\cite{haloscope4}. With an IF (Intermediate Frequency) of 38 MHz,
we take power spectra over a bandwidth of 3 MHz, which allows 10 power
spectra to overlap in most of the cavity resonant frequencies with a
discrete frequency step of 300 kHz. Power spectra are divided by the
noise power estimated from the measured and calibrated system noise
temperatures. The five-parameter fit also developed in
Ref.~\cite{haloscope4} is then employed to eliminate the residual
structure of the power spectrum.
The background-subtracted power spectra are combined in order to
further reduce the power fluctuation. We found no significant excess
from the combined power spectrum and thus set exclusion limits of
$g_{a\gamma\gamma}$ for $24.7<m_a<29.1$ $\mu$eV. No frequency bins in
the combined power spectrum exceeded a threshold of 5.5$\sigma_{P_n}$,
where $\sigma_{P_n}$ is the rms of the noise power $P_n$.
We found $\sigma_{P_n}$ was underestimated due to the five-parameter
fit as reported in Ref.~\cite{HF} and thus corrected for it
accordingly before applying the threshold of 5.5$\sigma_{P_n}$. Our
SNR in each frequency bin in the combined power spectrum was also
combined with weighting according to the Lorentzian lineshape,
depending on the $Q_L$ at each resonant frequency of the cavity.
With the tail of the assumed Maxwellian axion signal shape, the best
SNR is achieved by taking about 80\% of the signal and associate noise
power; however, doing so inevitably degrades SNR in Eq.~(\ref{EQ:SNR})
by about 20\%. Because the axion mass is unknown, we are also unable
to locate the axion signal in the right frequency bin, or
equivalently, the axion signal can be split into two adjacent
frequency bins.
On average, the signal power reduction due to the frequency
binning is about 20\%. The five-parameter fit also degrades the signal
power by about 20\%, as reported in Refs.~\cite{haloscope4, HF}. Taking
into account the signal power reductions described above, our SNR
for $g_{a\gamma\gamma}\sim6.5\times10^{-8}$ GeV$^{-1}$ is greater or
equal to 10, as mentioned earlier. The 95\% upper limits of the power
excess in the combined power spectrum are calculated in units of
$\sigma_{P_n}$; then, the 95\% exclusion limits of $g_{a\gamma\gamma}$
are extracted using $g_{a\gamma\gamma}\sim6.5\times10^{-8}$ GeV$^{-1}$
and the associated SNRs we achieved in this work. Figure~\ref{FIG:LIMIT}
shows the excluded parameter space at a 95\% confidence level (C.L.)
by the simplified ACTION experiment. We demonstrate an axion haloscope
with toroidal geometry in this paper and our result supersedes the
pre-existing exclusion limits reported by ALPS~\cite{ALPS2010} in the
relevant mass ranges.

\begin{figure}[htbp]
  \centering
  \includegraphics[width=0.52\textwidth,height=0.35\textwidth]{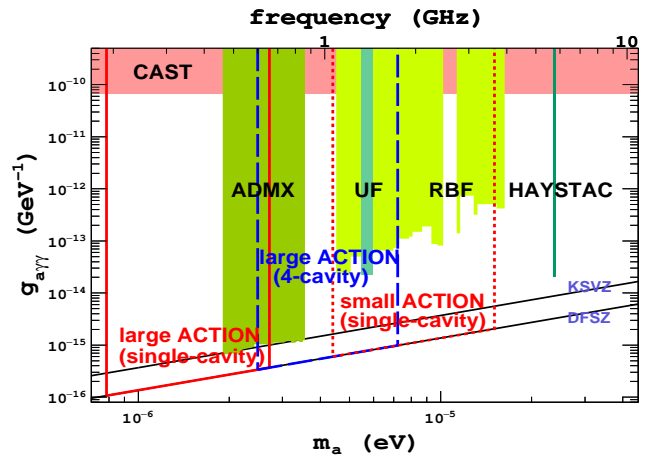}
  \caption{Expected exclusion limits by the large (solid lines with a
    single-cavity and dashed lines with a 4-cavity) and small (dotted
    lines with a single-cavity) ACTION experiments. Present exclusion
    limits from RBF~\cite{haloscope2}, UF~\cite{haloscope3},
    ADMX~\cite{haloscope4, haloscope5}, HAYSTAC~\cite{HF}, and
    CAST~\cite{CAST2017} are also shown.}
  \label{FIG:PROSPECTS}
\end{figure}

The prospects for axion dark matter searches with two larger-scale
toroidal geometries that could be sensitive to the KSVZ and DFSZ
models are now discussed. A similar discussion can be found
elsewhere~\cite{DIPOLE}.
One is called the ``large ACTION'', and the other is the ``small
ACTION'', where the cavity volume of the former is about 9,870 L and
that of the latter is about 80 L. The $B_{\rm avg}$ targets for the
large and small ACTION experiments are 5 and 12 T, respectively, where
the peak fields of the former and latter would be about 9 and 17
T. Hence, the large and small toroidal magnets can be realized by
employing NbTi and Nb$_3$Sn superconducting wires, respectively. The
details of the expected experimental parameters for the ACTION
experiments are listed in Table~\ref{TABLE:PROSPECTS}.
With an alumina tuning hoop whose relative permittivity is 9.9 and
QTM$_1$ modes, the large ACTION could search for axion dark matter
down to 0.79 $\mu$eV, and could search up to 2.73 $\mu$eV with a
copper tuning hoop and quasi-TM$_{020}$ (QTM$_2$) modes; hence, the
large ACTION could search for the axion mass range from 0.79 to 2.73
$\mu$eV with several configurations of tuning hoops and search
modes. With the same approach employed for the large ACTION, the
feasible search range of the small ACTION is also estimated.
The search range that would be very difficult to investigate in the
large and small ACTION experiments with a single-cavity could be
scanned by adopting a multiple-cavity~\cite{SWYOUN}, in this case, by
using the 4-cavity together with the large toroidal magnet.
The $Q_L$ values with $\beta=2$~\cite{OVERCOUPLE} were estimated with
pure copper and loss free alumina at cryogenic temperatures, but they
were limited by the anomalous skin effect.
Thanks to the very large volume of the cavity in the large ACTION
experiment, we could realize a scanning rate ($d\nu/dt$) that achieved
DFSZ sensitivity over a relevant search range after about 1 year of
scanning with state-of-the-art commercial transistor-based amplifiers
as the 1st amplifiers whose noise temperatures are about
1 K~\cite{LNF} and the physical temperature of the cavity
$T_{p,{\rm cavity}}$ of 4.2 K.
This temperature is, thus, equivalent to $T_{n,{\rm cavity}}\sim3.7$
K~\cite{T_CAVITY} which can be achieved easily
with liquid helium. On top of a $T_{p,{\rm cavity}}$ of 0.1 K,
quantum-noise-limited superconducting amplifiers were employed as the
1st amplifiers whose noise temperatures are proportional to not only
the frequencies ($T_{n,{\rm amp}}\sim h\nu/k_{B}\ln2$)~\cite{QN} but
also the ambient temperatures ($T_{p,{\rm cavity}}$ in axion
haloscopes) to achieve fast DFSZ scanning rates from the small ACTION
experiment. Figure~\ref{FIG:PROSPECTS} shows the exclusion limits
expected from the large and small ACTION experiments, and therein the
exclusion limits from RBF~\cite{haloscope2}, UF~\cite{haloscope3},
ADMX~\cite{haloscope4, haloscope5}, HAYSTAC~\cite{HF}, and
CAST~\cite{CAST2017} are also shown.

In summary, we have reported an axion haloscope search employing toroidal
geometry using the simplified ACTION experiment. The simplified ACTION
experiment excludes the axion-photon coupling $g_{a\gamma\gamma}$ down
to about $5\times10^{-8}$ GeV$^{-1}$ over the axion mass range from
24.7 to 29.1 $\mu$eV at the 95\% C.L. This is the first axion
haloscope search utilizing toroidal geometry since the advent of the
axion haloscope search by Sikivie. We have also discussed the
prospects for axion dark matter searches with two larger-scale
toroidal geometries, the large ACTION and small ACTION, that could be
sensitive to cosmologically relevant couplings over the axion mass
range from 0.79 to 15.05 $\mu$eV with several configurations of tuning
hoops, search modes, and multiple-cavity system.

\acknowledgments
This work was supported by IBS-R017-D1-2017-a00. B. R. Ko and J. Choi
acknowledge E. Won for his idea in winding a larger number of coils.


\begin{thebibliography}{99}
\bibitem{WHYNOT8GHz}
  Though we can tune the frequency from 6 to 8 GHz with the cavity and
  the frequency tuning system, the couplings are too weak with the
  resonant frequencies higher than 7 GHz. Therefore, we limit the
  range up to 7 GHz, which does not affect the purpose and
  significance of this work so much. We also found that the couplings
  can be improved by optimizing the magnetic loop coupling positions.
\bibitem{T_CAVITY}
  $T_{n,\rm cavity}$ and $T_{p,\rm cavity}$ are connected with $\beta$;
  $T_{n,\rm
  cavity}=\big(1-\big(\frac{\beta-1}{\beta+1}\big)^2\big)T_{p,{\rm
    cavity}}$ at the resonant frequency, thus when $\beta=1$ (or a
  cavity is coupled critically), $T_{n,\rm cavity}=T_{p,\rm cavity}$.
\end{thebibliography}
\end{document}